\documentclass[aps,prl,fleqn,floatfix,twocolumn,superscriptaddress,twoside,showpacs,footinbib]{revtex4}

\usepackage[dvipdfm]{graphicx}

\usepackage{amsmath,amssymb,amsfonts}

\usepackage{color}
\usepackage{eucal}

\newcommand \ra  {\rightarrow}

\newcommand \vq {\vec{q}}

\newcommand \A {\alpha}

\newcommand \D {\Delta}

\newcommand \lc {\langle}
\newcommand \rc {\rangle}

\newcommand \prt {\partial}

\newcommand \bvec{\left( \begin{array}{c} }
\newcommand \evec{\end{array} \right)}

\newcommand \bea{\begin{eqnarray} }
\newcommand \eea{\end{eqnarray} }
\newcommand \nn {\nonumber}
\newcommand \be {\begin{equation}}
\newcommand \ee {\end{equation}}

\newcommand \ata {& \times &}

\begin{document}

\title{A pQCD-based description of heavy and light flavor jet quenching }
\author{Guang-You Qin}
\affiliation{Department of Physics, Duke University, Durham, NC 27708, USA}
\affiliation{Department of Physics, The Ohio State University, Columbus, OH, 43210, USA}
\author{Abhijit Majumder}
\affiliation{Department of Physics, The Ohio State University, Columbus, OH, 43210, USA}

\date{\today}
\begin{abstract}

We present a successful description of the medium modification of light and heavy flavor jets 
within a perturbative QCD (pQCD) based approach.
Only the couplings involving hard partons are assumed to be weak. 
The effect of the medium on a hard parton, per unit time, is encoded in terms of three non-perturbative, related  transport coefficients which describe  
the transverse momentum squared gained, the elastic energy loss and diffusion in elastic energy transfer. 
Scaling the transport coefficients with the temperature of the medium, we achieve a good description of the centrality 
dependence of the suppression and the azimuthal anisotropy of leading hadrons. Imposing additional constraints based on 
leading order (LO) Hard Thermal Loop (HTL) effective theory leads to a worsening of the fit, implying the necessity of computing transport coefficients beyond LO-HTL. 

\end{abstract}
\maketitle

Experimental results from the Relativistic Heavy Ion Collider (RHIC) have established that 
a new kind of hot and dense partonic matter has been created in central Au+Au collisions~\cite{sqgp}. 
High transverse momentum ($p_T$) partons created in the initial hard collisions were 
predicted to provide a reliable probe of this highly 
excited matter~\cite{Gyulassy:1993hr}. These partons lose energy in the dense medium 
leading to a depleted yield of high $p_T$ hadrons compared to 
that in binary scaled p+p collisions~\cite{light_flavor}. 
Due to the large energy scale involved, these 
hard partons were expected to couple weakly with the medium allowing for the use of 
perturbative QCD (pQCD) to describe their propagation through the dense matter.
Experimentally, the basic picture of parton energy loss has been confirmed by the observation of 
significant suppression of the high $p_T$ yield for both light~\cite{light_flavor} and heavy flavors~\cite{heavy_flavor}. 
There now exist sophisticated (and successful) calculations of light flavor suppression~\cite{Bass:2008rv}.
Not only have these accounted for the centrality dependence and azimuthal anisotropy of the single inclusive 
suppression but also for photon and hadron triggered correlations on both the near~\cite{Majumder:2004pt} and away side of the trigger hadron~\cite{photon_hadron_corr}.

The success of pQCD based calculations of heavy flavor modification, however, has been less than satisfactory.
In a prior effort, 
Armesto~\emph{et al.}~\cite{Armesto:2005mz} were able to describe both light and heavy flavor suppression in 
central collisions (and the azimuthal anisotropy) 
including only radiative energy loss. 
However, they required a time averaged jet transport parameter $\hat{q} \equiv d(\Delta p_\perp)^2/dt \sim 14$ GeV$^{2}/$fm for a gluon jet 
($p_\perp$ is the momentum trasverse to the jet axis and $t$ is the time spent in the medium), yielding a $p_\perp$ comparable to the energy of the 
parent jet and at least a factor of five larger than estimates from LO-HTL using a medium with an identical temperature profile.
In a recent analysis by 
Wicks~\emph{et al.}~\cite{Wicks:2005gt}, the authors incorporated both 
radiative and elastic energy loss. The elastic energy loss coefficient $\hat{e} = dE/dt$ 
was calculated in LO-HTL but the radiative energy loss was calculated in a different medium of static 
scattering centers, estimating the density of scatterers and the transverse 
momentum per scattering with the help of LO-HTL calculations constrained by entropy considerations. 
These authors were unable to \emph{quantitatively} describe the suppression of non-photonic electrons 
in central collisions at RHIC. This has led the authors of Ref.~\cite{Horowitz:2007su} 
to question the ability of any pQCD based approach to consistently describe 
high $p_T$ suppression, leading to the speculation that hard partons with 
energies in the tens of GeV may be strongly interacting with the produced dense medium~\cite{s_drag}. 
This has cast a pall of doubt on the entire program of pQCD based jet tomography of dense 
strongly interacting media.

In this Letter, we provide a satisfactory explanation of 
both light and heavy flavor suppression (including its centrality dependence and azimuthal anisotropy) 
within a single pQCD based approach. Taken in combination with 
the results of Refs.~\cite{Bass:2008rv,Majumder:2004pt,photon_hadron_corr}, this effort closes 
the gap in the suite of hard probe observables at RHIC that can be successfully described by pQCD.
We base our approach on the factorization paradigm inherent to the higher twist scheme, 
where a hard jet is weakly coupled with 
the gluon distribution of the medium, which itself may be strongly or weakly coupled. 
The effect of the medium is encoded in terms of three non-perturbative transport coefficients: 
$\hat{q}$, $\hat{e}$, and the diffusion in elastic energy transfer per unit time $\hat{e}_2 = d(\Delta E)^2/dt$.

The primary difference with Ref.~\cite{Wicks:2005gt} is that we do not insist on estimating the transport 
coefficients in LO-HTL but instead treat them as parameters of the model.
Assuming near on-shell propagation for the hard partons, 
we relate $\hat{e}$ and $\hat{e}_2$ to the loss in 
longitudinal momentum per unit time (${dp_z}/{dt}$) and diffusion per unit time in 
longitudinal momentum transfer (${d (\D p_z)^2 }/{dt}$). 
Assuming a medium close to local thermal equilibrium and the applicability of the fluctuation-dissipation theorem~\cite{Moore:2004tg}, 
we set ${d (\D p_\perp)^2}/{dt} \simeq 2 { d (\D p_z)^2 }/{dt} \simeq ({4T}/{|v|}){dp_z}/{dt}$, with $v$ the velocity of the jet parton. 
The only input parameter is chosen as $\hat{q}$ (assuming the same for both light and heavy quarks) which is assumed to scale with temperature ($\hat{q} = C T^3$).  
This form is similar to the results obtained for a hard jet weakly coupled to a strongly coupled 
medium~\cite{Liu:2006ug} and also suggested by higher order calculations of $\hat{q}$ in HTL effective 
theory~\cite{CaronHuot:2009zz}. With such setting, we achieve a satisfactory description of the medium modification of both light and heavy 
flavor jets and their elliptic flows.
For completeness, we also perform a fit with the form of the coefficients taken from LO-HTL theory 
(with the in-medium $\alpha_s$ as the fit parameter). While the transport coefficients obtained from the fit turn out to be comparable, 
a worsening of the fit is obtained, in agreement with Ref.~\cite{Wicks:2005gt}. 
Thus, while the applicability of LO-HTL effective theory to the calculation of medium properties may be questioned, 
the assumption that a hard jet is weakly coupled with the medium, allowing for a pQCD based description of light and heavy 
flavor suppression, is further verified and established.

High momentum partons produced in hard collisions, tend to be rather virtual. 
In vacuum, virtuality ($Q^2$) is lost by subsequent 
emissions. The probability for these emissions can be calculated using 
pQCD as long as the virtuality at a given emission is large enough.
Below a certain  $Q^2$, one will need to use an experimentally fitted fragmentation~function~(FF) to calculate the 
inclusive distribution of single hadrons. The change of this FF due to prior higher virtuality 
(transverse momentum) emissions can be calculated using the Dokshitzer-Gribov-Lipatov-Altarelli-Parisi\ (DGLAP) evolution equations~\cite{DGLAP}.
These require the measured FF at some lower scale $\mu^2$ as input and 
compute the modification to this single hadron distribution due to multiple emissions from 
$\mu^2$ up to $Q^2$.

Hard virtual partons traversing a dense medium will scatter off the constituents of the medium in 
addition to emitting lower virtuality partons; this changes the momentum distribution of the propagating 
partons between emissions. 
A parton with light-cone momentum $q^-\!\!$ ($q^-\!\!= \! q^0\!\! - q_z$, for a parton traveling in the $-z$ direction)  
much larger than its virtuality $q^- \!\!\gg Q$, can be effectively described with a length ($L^-$) dependent 
three dimensional distribution $\phi(q^-,\vq_\perp,L^-)$. 
The change in this distribution due to multiple scattering per unit length, up to second order in gradients,  
can be expressed as~\cite{ehat_qhat}
\begin{eqnarray}
\frac{\prt \phi}{\prt L^- } &=&  \hat{q}_{\rm lc} \nabla_{q_\perp}^2 \phi 
+ \hat{e}_{\rm lc}\frac{\prt \phi}{\prt q^-} + {\hat{e}_{2 \rm lc}} \frac{\prt^2 \phi}{\prt {q^-}^2} ,
\end{eqnarray}
where, $\hat{q}_{\rm lc}$, $\hat{e}_{\rm lc}$ and $\hat{e}_{2 \rm lc}$ 
are non-perturbative transport coefficients which encode properties of  the medium. 
These can be expressed in terms of two gluon operators, e.g., 
\begin{eqnarray} \label{qhat_ehat_def}
\hat{q}_{\rm lc} \!\!\!&=&\!\!\! \left[{4 \pi^2 \A_s C_R}/(N_c^2 - 1)\right] \!\!\!\int\!\! d y^- \!\lc F^{+ \mu} (y^-) F_{\mu}^+ (0)\rc,  \\
\hat{e}_{2 \rm lc}\!\!\! &=& \!\!\!\left[{4 \pi^2 \A_s C_R}/{(N_c^2 - 1)}\right] \!\!\!\int \!\!dy^- \!\lc F^{+ -} (y^-) F^{+-} (0) \rc. \,\,\,\,\,\,\,\,\,\,\,\,\,\, \mbox{}\nn 
\end{eqnarray}
These operator products can be factorized from the hard process and computed in any given model of the medium. 
The angled brackets $\lc \rc$ indicate an expectation of the operator product in an arbitrary ensemble.
We do not present the operator expression for the elastic loss $\hat{e}_{\rm lc}$ as we will always relate it 
to the expression for fluctuations in the elastic loss, $\hat{e}_{2 \rm lc}$, via the fluctuation-dissipation theorem. 
These light-cone transport coefficients may be related to the Cartesian coefficients as
\begin{eqnarray}
\hat{q} = (1+|v|)\hat{q}_{\rm lc} \,, \hat{e} = |v|\hat{e}_{\rm lc} \,,  \hat{e}_2 = v^2/(1+|v|)\hat{e}_{2 \rm lc} \,. \ \ \ 
\end{eqnarray}

The multiple scattering and emissions from a hard virtual parton and its effect on the 
final single hadron distribution can be calculated as long as the virtualities, both on entry and exit from the 
medium, are large enough for pQCD to be applicable. For the case of only transverse scattering,  
when $Q^2 \gg k_\perp^2$,  
the soft scale of the medium, one recovers a DGLAP like picture with ordered emissions given by a 
splitting function which is modified by multiple transverse scattering in the medium~\cite{Majumder:2009zu}. 
The medium modified fragmentation 
function (MMFF), for a hadron with momentum fraction $z$ of the jet, now depends on the location of entry ($\zeta_i$) and exit ($\zeta_f$) from the medium; its 
change with virtuality is given as, 
\begin{eqnarray}\label{MMFF}
 \frac{\partial \tilde{D}_i(z, Q^2,q^-)|_{\zeta_i}^{\zeta_f}}{\partial \ln Q^2} \!\!&=&\!\! \!\sum_j\! 
 \frac{\alpha_s}{2\pi} \!\int_z^1\!\! \frac{dy}{y} \!\int_{\zeta_i}^{\zeta_f} \!\!\!\! d\zeta \tilde{P}_{i\to j}(y,\zeta,Q^2,q^-) \nn \\ 
 \!\!\ata\!\!  \tilde{D}_j({z}/{y},Q^2,q^-y)|_{\zeta}^{\zeta_f}.  \label{mod_D} \ \ \ \
\end{eqnarray}
In the equation above, $\tilde{P}_{i \to j}$ is the in-medium splitting function (IMSF) at location $\zeta$~\cite{Majumder:2009zu},
\begin{eqnarray}\label{IMSF}
\tilde{P}_{i\to j} = \frac{P_{i \to j}(y)}{Q^2} \frac{\hat{q}(\zeta)}{\pi} 
\left[2-2\cos\left(\frac{\zeta-\zeta_i}{\tau_{\rm f}}\right)\right], \
\end{eqnarray}
where $\tau_{\rm f}$ is the formation time of the emitted gluon with forward 
momentum fraction $y$ and transverse momentum $l_\perp=Q$:
$\tau_{\rm f} = 2q^-y(1-y) / l_\perp^2$.
The full evolution equations for MMFFs will also include a pure vacuum contribution, which is implicitly included.
By taking mass effects into account, the above results have also been extended to heavy 
quarks~\cite{Zhang:2004qm}, which suffer less radiative energy loss due to the 
dead cone effect.
Here $\tau_{\rm f} = 2q^-y(1-y)/[l_\perp^2 + (1-y)^2 M_Q^2]$, 
with $M_Q$ the heavy quark mass. The IMSF also contains a multiplicative factor $Q^8/[Q^2+(1-y)^2M_Q^2]^4$.

In this calculation, the masses of the heavy quarks and heavy mesons are taken 
as: $M_c = 1.3$~GeV, $M_b = 4.2$~GeV, $M_D = 1.9$~GeV and $M_B = 5.3$~GeV.
Our input to the evolution equations is picked using restrictions based on formation length: 
given a mean length traversed of $\lc L^- \rc$, 
partons with formation lengths ($\tau = q^-/Q^2$) much larger than this will not be modified by the medium, as a 
result, we take as input the vacuum fragmentation function at $\mu_0 = q^-/L^-$. 
In cases where $\mu_{0}^{2}$ falls below $1$ GeV$^{2}$, $\mu_{0}^{2}$ is set equal to 
$1$ GeV$^{2}$. These fragmentation functions are then evolved up to the hard scale using 
a sum of the medium modified evolution equations [Eq.~\eqref{mod_D}] and the standard 
vacuum evolution equations. 
The modification due to elastic energy loss is incorporated by shifting the fraction $z$,
\begin{eqnarray}\label{Dprime}
{D}'(z) = \int d\Delta z P(\Delta z) {{D}\left({z}/({1-\Delta z})\right)}/({1-\Delta z}), \
\end{eqnarray}
where $P(\Delta z)$ is a Gaussian distribution with a mean and variance 
determined by $\hat{e}$ and $\hat{e}_2$ respectively.

Final hadron spectra are obtained as the convolution,
\begin{eqnarray}\label{dsigmah}
 {d\sigma_h} = \sum_{abd}{f_{a/A} \otimes f_{b/B} \otimes d\sigma_{ab\to jd}} \otimes \tilde{D}_{h/j},
\end{eqnarray}
where $f_{(a,b)/(A,B)}$ is the nuclear parton distribution function (PDF), $d\sigma$ is the partonic cross section, and $\tilde{D}_{h/j}$ 
is the FF after radiative and collisional energy losses. 
For both light and heavy sectors, PDFs are taken from CTEQ5 \cite{Lai:1999wy} 
(shadowing corrections are obtained from EKS98~\cite{Eskola:1998df}). The partonic cross section $d\sigma$ is evaluated 
at leading order, with a $K$-factor accounting for next-to-leading order effects ($1.7$ and $2.4$ for light and heavy sectors, repectively). 
The renormalization and 
factorization scales are set to be the transverse energy of hadrons $E_T^h=({p_T^h+M_h^2})^{1/2}$. The vacuum 
FFs for light quarks and gluons are taken from KKP \cite{Kniehl:2000fe} and $c$$\ra$$D$, $b$$\ra$$B$ 
FFs  are obtained from PYTHIA6.4 \cite{Sjostrand:2006za}.
We introduce multiplicative semi-leptonic decay functions of heavy mesons  ($f_{e/H}$) to obtain the non-photonic electron spectrum.
They are obtained by fitting to $D$ decay from 
BABAR \cite{Aubert:2004td} and $B$ decay from CLEO \cite{Adam:2006nu}. 
The normalizations are determined from their branching ratios: 
${\rm BR}(D$$\to$$e) \approx 10\%$ and ${\rm BR}(B$$\to$$e) = 10.36\%$ \cite{Adam:2006nu}. 
With these, we obtain a reasonable description of the baseline experimental measurements of $\pi^0$ spectra \cite{Adler:2003pb} 
and non-photonic electron production \cite{Adare:2006hc} in p+p collisions.

Given the dependence of $\hat{q},\hat{e}$ and $\hat{e}_2$ on temperature $T$, 
a midrapidity thermal space-time profile $T(\vec{r}_\perp, \tau)$ is required. 
In this application, the initial transverse spatial profile 
for the entropy density $s\sim T^3$ 
is chosen to be proportional to the participant density in the collision of 
two nuclei with a Woods-Saxon nuclear density function (parameters are taken from Ref.~\cite{De Jager:1974dg}).
The medium is assumed to thermalize at proper time 
$\tau_0 = 0.6$~fm/c, with a temperature $T_0 = 400$~MeV at the hottest point in central collisions. 
The temperature diminishes with time as $\tau^{-1/3}$ due to one dimensional Bjorken expansion.
The spatial distribution of the initial hard jets is determined by the binary nucleon collision density.

\begin{figure}[htbp]
\includegraphics[width=0.8\linewidth]{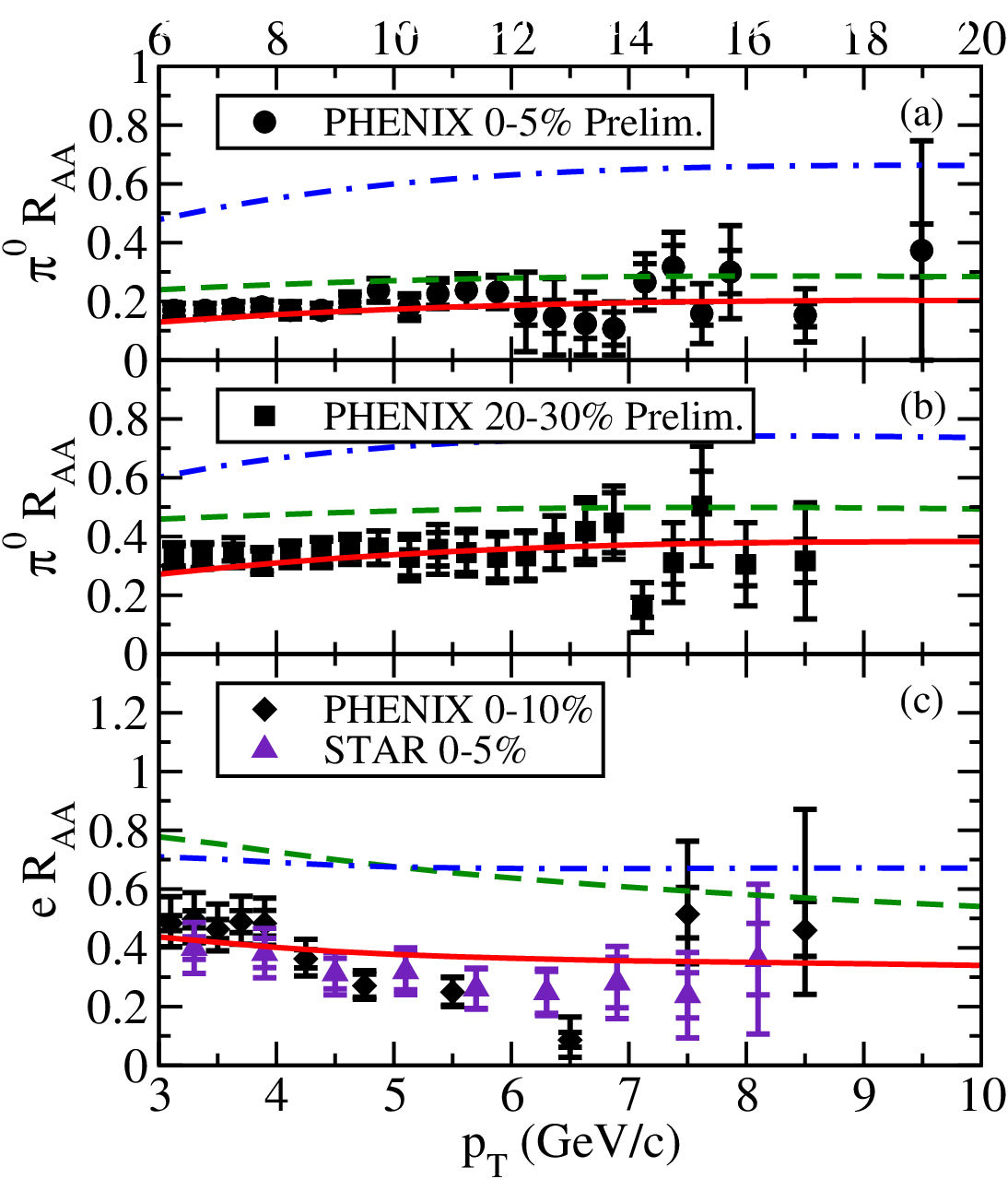} 
 \caption{(Color online) $R_{AA}$ (solid lines) for $\pi^0$  
in central (a) and mid-central (b)   
and non-photonic electrons in central (c) Au+Au collisions at RHIC, 
with $\hat{q} \propto T^3$ (and corresponding $\hat{e}$ and $\hat{e}_2$). 
Dashed and dot-dashed lines represent the radiative and collisional contributions, respectively. 
} \label{raa_fig}
\end{figure}
\begin{figure}[htbp]
\includegraphics[width=0.9\linewidth]{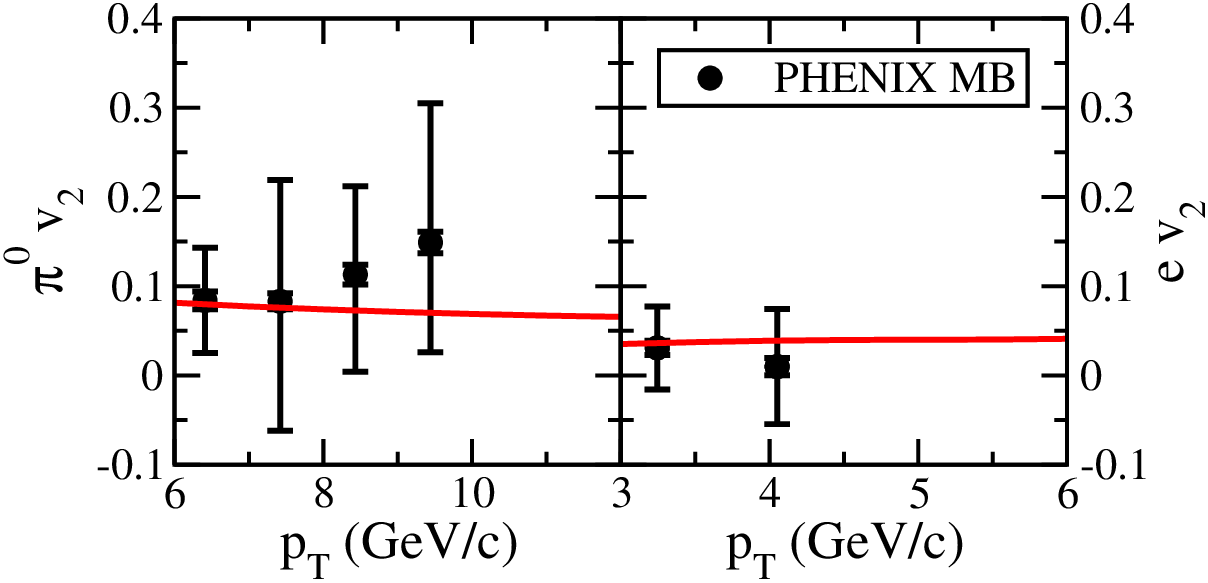} 
 \caption{(Color online) The predicted minimum bias $v_2$ for $\pi^0$ and heavy 
flavor electrons in Au+Au collisions at RHIC, using the same parameters as determined in Fig. \ref{raa_fig}.
 } \label{v2}
\end{figure}

We compare with the measured nuclear modification factor ($R_{AA}$) 
as a function of $p_T$, defined as the ratio of the yield in $A+A$ to the binary scaled yield in $p+p$, 
\begin{eqnarray}\label{raa}
R_{AA} = \frac{1}{N_{\rm coll}}\frac{dN^{AA}/d^2p_Tdy}{dN^{pp}/d^2p_Tdy},
\end{eqnarray}
where $N_{\rm coll}$ is the average number of binary collisions between the nucleons from two nuclei in the range of impact parameters chosen. 
With a single $\hat{q}_0 \approx 1.3$GeV$^2$/fm (and associated values of $\hat{e}, \hat{e}_2$) for a quark jet at $T = 400$~MeV, 
a satisfactory description of $R_{AA}$ for $\pi^0$ in central and semi-central collisions and the supression of heavy-flavor electrons in 
central collisions is achieved [see Fig.~\ref{raa_fig}]. It is clear that the inclusion of elastic energy loss \cite{Qin:2007rn} is necessary 
to explain the suppression of both light and heavy flavor suppressions.
Using these values we can predict the impact parameter integrated azimuthal anisotropy 
[minimum bias $v_2$, (Fig.~\ref{v2})].
A non-Glauber-based initial profile  will affect this final predicted $v_2$, as will a more realistic calculation in a three dimensional 
hydrodynamic simulation. 

\begin{figure}[htbp]
\includegraphics[width=0.8\linewidth]{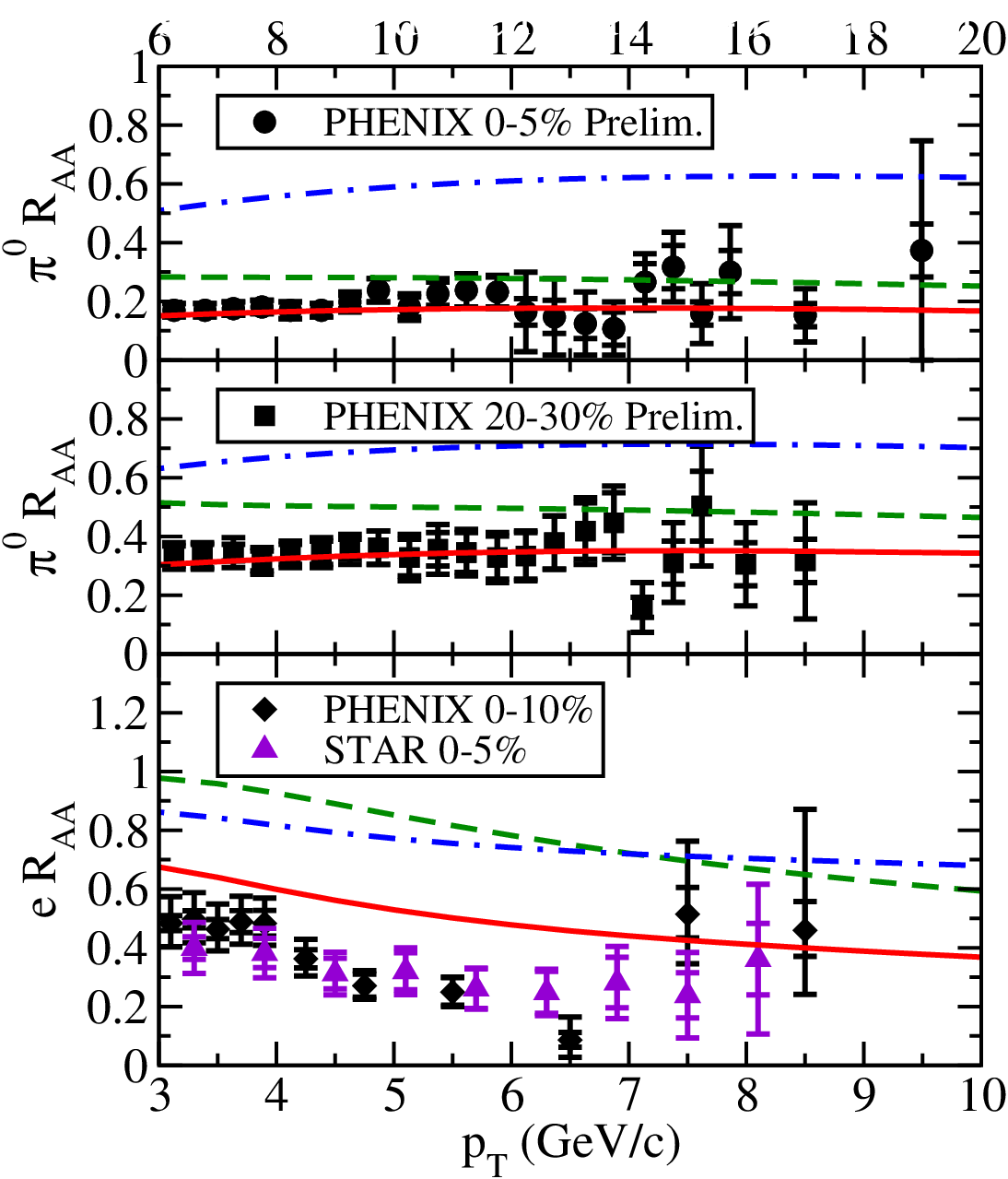} 
\caption{(Color online) 
Same as Fig.~\ref{raa_fig}, but with 
$\hat{q}, \hat{e}$ and $\hat{e}_2$ calculated in LO-HTL. } \label{raa_htl}
\end{figure}

We close with a discussion of a weakly coupled medium where the transport coefficients are estimated using LO-HTL theory. This gives, for a light quark or gluon, 
\begin{eqnarray} \label{eqhat_htl}
\hat{q}_{\rm HTL} = C_R \alpha_s m_D^2 T \ln \left[{4ET}/{m_D^2}\right], 
\end{eqnarray}
where $m_D^2 = 4\pi\alpha_s (1+N_f/6) T^2$ is the Debye mass ($N_f=3$). LO-HTL expressions for the heavy quark $\hat{e}$ are obtained from Ref.~\cite{Braaten:1991we}. 
The in-medium coupling $\alpha_s$ is adjusted to describe the central collision data for light flavors; this yields $\hat{q}_0 \approx 1.3$~GeV$^2$/fm for a quark jet 
with $E=20$~GeV. Using these we predict the results for non-central collisions and heavy quark energy loss. 
In agreement with Ref.~\cite{Wicks:2005gt}, we find (see Fig.~\ref{raa_htl}) a considerable worsening of the comparison with the 
heavy flavor data ($\chi^2/{\rm d.o.f.} = 110/20 \approx 5.5$, compared with $\chi^2/{\rm d.o.f.} = 24/20 \approx 1.2$ in Fig. \ref{raa_fig} ).

In conclusion, we have carried out a global fit of high $p_T$ single inclusive observables 
for both light and heavy flavors at RHIC, where only 
the coupling of the jet with the medium is assumed to be weak. The properties of the medium are encoded by non-perturbative transport coefficients $\hat{q}$, 
$\hat{e}$ and $\hat{e}_2$, which are related by the fluctuation dissipation theorem for an isotropic thermal medium.  
With only one input parameter (chosen as $\hat{q}$) scaling with $T^3$, we achieve a good agreement with the experimental measurements for both 
the suppression and the elliptic flow.
This supports the assertion that perturbative QCD and weak coupling approaches can be applied 
to the description of jet modification in a dense medium, even if the medium may itself not be weakly coupled. 
Insisting on estimating all the transport coefficients solely within LO-HTL theory leads to a failure to explain the heavy flavor suppression, implying the need to 
go beyond LO-HTL to compute jet transport coefficients.

We thank Ulrich Heinz and Berndt M\"{u}ller for helpful discussions. This work was supported in part by U.S. D.~O.~E. under grant Nos. DE-FG02-01ER41190 and DE-FG02-05ER41367.
\vspace{-0.0cm}

\vspace{-0.5cm}



\end{document}